\documentclass{webofc}
\usepackage[varg]{txfonts}   % Web of Conferences font
\usepackage{graphicx}  % Include figure files
\usepackage{amsmath}
\usepackage{physics}
\usepackage[dvipsnames]{xcolor}
\usepackage{fancyhdr}
\begin{document}

\title{Numerical Convergence of Electromagnetic Responses with the Finite-Amplitude Method}

\author{
\firstname{Tong} \lastname{Li}\inst{1}\fnsep\thanks{\email{li94@llnl.gov}} 
\and
\firstname{Nicolas} \lastname{Schunck}\inst{1}\fnsep\thanks{\email{schunck1@llnl.gov}} 
}

\institute{Nuclear and Chemical Sciences Division, 
Lawrence Livermore National Laboratory, Livermore, California 94551, USA}

\abstract{The response of a nucleus to an electromagnetic probe is a key 
quantity to simulate photabsorption or photodeexcitation processes. For large 
calculations at the scale of the entire mass table, this response can be 
estimated by linear response theory. Thanks to the introduction of the 
finite-amplitude method (FAM), calculations are computationally efficient. In this 
paper, we investigate in more details the convergence of FAM calculations of 
the response function as a function of the parameters controlling the numerical 
implementation of the theory. We show that the response is much less sensitive 
to the details of the single-particle basis than, e.g., Hartree-Fock-Bogoliubov
calculations.
}
\maketitle
\thispagestyle{fancy}
%

%%%%%%%%%%%%%%%%%%%%%%%%%%%%%%%%%%%%%%%%%%%%%%%%%%%%%%%%%%%%%%%%%%%%%%%%%%%%%%%
%%%%%%%%%%%%%%%%%%%%%%%%%%%%%%%%%%%%%%%%%%%%%%%%%%%%%%%%%%%%%%%%%%%%%%%%%%%%%%%
%%%%%%%%%%%%%%%%%%%%%%%%%%%%%%%%%%%%%%%%%%%%%%%%%%%%%%%%%%%%%%%%%%%%%%%%%%%%%%%
%%%%%%%%%%%%%%%%%%%%%%%%%%%%%%%%%%%%%%%%%%%%%%%%%%%%%%%%%%%%%%%%%%%%%%%%%%%%%%%

\section{Introduction}
\label{sec:intro}

Photonuclear processes play central roles as probes to nuclear structure and 
in applications of nuclear physics to nuclear technology, isotope 
production or astrophysics. Much work is thus devoted to building reliable 
and comprehensive data libraries of properties such as $\gamma$-strength 
functions \cite{goriely2019reference}. Notwithstanding fully phenomenological 
approaches based on explicit fit to experimental data, large-scale theoretical 
predictions of such quantities typically rely on the quasiparticle random-phase
approximation (QRPA) and linear response theory \cite{goriely2002largescale,
goriely2004microscopic,daoutidis2012largescale,martini2016largescale,
goriely2018gognyhfb}. Because of their unfavorable computational scaling, most 
of these calculations have often been restricted to spherical symmetry and/or 
even-even nuclei.

With the introduction of the finite-amplitude method (FAM), the computational 
cost of the linear response in deformed nuclei became considerably lower and is 
now somewhat comparable to the one of a deformed Hartree-Fock-Bogoliubov (HFB) 
calculation \cite{nakatsukasa2007finite,avogadro2011finite}. This has enabled
studies of nuclear multipole responses in heavy, deformed 
nuclei \cite{inakura2009selfconsistent,stoitsov2011monopole,
kortelainen2015multipole,oishi2016finite} as well as more accurate estimates of 
nuclear moments of inertia and the collective inertia tensor 
\cite{petrik2018thoulessvalatin,washiyama2021finiteamplitude}. The FAM was 
originally introduced and tested with non-relativistic Skyrme functionals and later also implemented with 
relativistic functionals \cite{niksic2013implementation}. 

In most of these early applications, the uncertainties of the results induced by 
numerical inputs such as the basis parameters, quasiparticle cutoff and 
smearing width were not fully discussed, with the exception of Refs.
\cite{stoitsov2011monopole} and \cite{oishi2016finite} which presented results 
for two different basis sizes. Such uncertainties should be quantified before 
embarking on any large-scale calculations of the response function. The goal of 
this paper is thus to perform a more systematical study of the convergence of 
FAM calculations with Skyrme functionals as a function of various numerical 
features of the implementation. We briefly recall the theoretical formalism in 
Section \ref{sec:theory} before presenting a selection of the results in 
Section \ref{sec:results}.

%%%%%%%%%%%%%%%%%%%%%%%%%%%%%%%%%%%%%%%%%%%%%%%%%%%%%%%%%%%%%%%%%%%%%%%%%%%%%%%
%%%%%%%%%%%%%%%%%%%%%%%%%%%%%%%%%%%%%%%%%%%%%%%%%%%%%%%%%%%%%%%%%%%%%%%%%%%%%%%
%%%%%%%%%%%%%%%%%%%%%%%%%%%%%%%%%%%%%%%%%%%%%%%%%%%%%%%%%%%%%%%%%%%%%%%%%%%%%%%
%%%%%%%%%%%%%%%%%%%%%%%%%%%%%%%%%%%%%%%%%%%%%%%%%%%%%%%%%%%%%%%%%%%%%%%%%%%%%%%

\section{Theoretical Framework}
\label{sec:theory}

We use the general framework of the energy density functional (EDF) to compute the 
electromagnetic response of nuclei. The theory is presented in great 
details in textbooks and review articles, most notably 
\cite{valatin1961generalized,mang1975selfconsistent,bender2003selfconsistent,
ring2004nuclear,schunck2019energy}. In the following, we only recall the most 
essential formulas.

%%%%%%%%%%%%%%%%%%%%%%%%%%%%%%%%%%%%%%%%%%%%%%%%%%%%%%%%%%%%%%%%%%%%%%%%%%%%%%%
%%%%%%%%%%%%%%%%%%%%%%%%%%%%%%%%%%%%%%%%%%%%%%%%%%%%%%%%%%%%%%%%%%%%%%%%%%%%%%%

\subsection{Hartree-Fock-Bogoliubov Theory with Skyrme Functionals}
\label{subsec:hfb}

The ground-state many-body wavefunction $\ket{\Phi}$ of the nucleus is computed 
at the Hartree-Fock-Bogoliubov (HFB) approximation. We recall that $\ket{\Phi}$ 
is then a product state of quasiparticle annihilation operators that are 
related to any single-particle basis by a (unitary) Bogoliubov, or canonical, 
transformation $\mathcal{W}$ characterized by the matrices $U$ and $V$,
\begin{align}
\beta_{\mu} & = \sum_{m} \qty[ U^{\dagger}_{\mu m}\,c_{m} + V^{\dagger}_{\mu m}\,c_{m}^{\dagger} ] \,,
\medskip\\
\beta_{\mu}^{\dagger} & = \sum_{m} \qty[ V^{T}_{\mu m}\,c_{m} + U^{T}_{\mu m}\,c_{m}^{\dagger} ]\, ,
\label{eq:bogoQP}
\end{align}
where $\qty{ \beta, \beta^{\dagger} }$ are the quasiparticle operators and 
$\qty {c, c^{\dagger} }$ the single-particle operators. The one-body density 
matrix $\rho = V^{*}V^{T}$ and pairing tensor $\kappa = V^{*}U^{T}$ can both be 
expressed as functions of the Bogoliubov transformation, and the total energy 
is a functional of $\rho$, $\kappa$ and $\kappa^{*}$. In practice, the 
matrices $U$ and $V$ are determined by requiring that the total energy 
$E[\rho,\kappa,\kappa^{*}]$ be a minimum with respect to all independent matrix 
elements $\rho_{ij}$, $\kappa_{ij}$, and $\kappa^{*}_{ij}$. The solution to the 
resulting HFB equation fully determines the Bogoliubov matrices $U$ and $V$.

In this paper, we choose the functional form $E[\rho,\kappa,\kappa^{*}]$ to be 
a Skyrme functional in the particle-hole channel (terms proportional to $\rho$) 
and a simple zero-range, density-dependent functional in the particle-particle 
channel (terms proportional to $\kappa$ and $\kappa^{*}$). As is well known, 
the zero range of the pairing functional requires applying a cutoff 
on the number of quasiparticles used to compute the density matrix and pairing tensors. % Tong: I want the first appearance of E_cut to be in Sec. "Quasiparticle cutoff".

%%%%%%%%%%%%%%%%%%%%%%%%%%%%%%%%%%%%%%%%%%%%%%%%%%%%%%%%%%%%%%%%%%%%%%%%%%%%%%%
%%%%%%%%%%%%%%%%%%%%%%%%%%%%%%%%%%%%%%%%%%%%%%%%%%%%%%%%%%%%%%%%%%%%%%%%%%%%%%%

\subsection{Linear Response Theory with the Finite-Amplitude Method}
\label{subsec:fam}

Linear response theory quantifies the effect of applying a small one-body 
perturbation operator $\hat{F}(t)$ to the ground state described by the 
many-body wavefunction $\ket{\Phi}$ and is most naturally derived as the 
small-amplitude limit of time-dependent Hartree-Fock-Bogoliubov theory (TDHFB)
\begin{equation}
i\hbar \pdv{\hat{\mathcal{R}}}{t} = \qty[ \hat{\mathcal{H}}(t) + \hat{F}(t), \hat{\mathcal{R}}(t) ]\, ,
\end{equation}
where $\hat{\mathcal{R}}(t)$ is the time-dependent generalized density and  
$\hat{\mathcal{H}}(t)$ the time-dependent HFB matrix. Here, one will assume 
that 
\begin{equation}
\hat{F}(t) = \hat{F}(\omega)e^{-i\omega t} + \hat{F}^{\dagger}(\omega)e^{+i\omega t} \,,
\end{equation}
where the frequency $\omega$ plays the role of the excitation energy in linear response theory. 
Under the effect of the perturbation operator, the generalized density 
$\hat{\mathcal{R}}_{0}$ associated with the HFB ground state $\ket{\Phi}$ 
becomes, to the first order, $\hat{\mathcal{R}}(t) = \hat{\mathcal{R}}_{0} 
+ \delta\hat{\mathcal{R}}(\omega)e^{-i\omega t} + \delta\hat{\mathcal{R}}(\omega)^\dagger e^{i\omega t}$, 
and correspondingly $\hat{\mathcal{H}}(t) = \hat{\mathcal{H}}_{0} 
+ \delta\hat{\mathcal{H}}(\omega)e^{-i\omega t} + \delta\hat{\mathcal{H}}(\omega)^\dagger e^{i\omega t}$.
In the Bogoliubov (quasiparticle) basis, the 
perturbed density $\delta\hat{\mathcal{R}}(\omega)$, perturbed mean field $\delta\hat{\mathcal{H}}$,  
and perturbation operator $\hat{F}(\omega)$ take the form
\begin{equation}
\delta\mathcal{R}(\omega) = 
\mqty(  0 & X \\
        Y^T & 0
     )\, ,
\qquad
\delta\mathcal{H}(\omega) = 
\mqty(  0 & \delta H^{20} \\
        -\delta H^{02} & 0
     )\, ,
\qquad
\mathcal{F}(\omega) = 
\mqty(  0 & F^{20} \\
      -F^{02} & 0
     )\, .
\end{equation}
The standard, matrix equation of linear response reads
\begin{equation}
\left[
\mqty(  A & B \\
        B^{*} & A^{*}
     )
-\omega
\mqty( 1 & 0 \\
       0 & -1
     )
\right]
\mqty( X(\omega) \\ Y(\omega) )
= 
\mqty( F^{20}(\omega) \\ F^{02}(\omega) )\, ,
\label{eq:LR_matrix}
\end{equation}
where the matrices $A$ and $B$ together form the QRPA matrix and are obtained 
from the second derivatives of the energy with respect to the generalized density $\mathcal{R}$. 
% the densities $\rho$, $\kappa$, and $\kappa^{*}$. 
% If the energy derives from a Hamiltonian $\hat{H}$, 
% these matrices can be also written as $A_{ij\mu\nu} = 
% \ev{\beta_{j}\beta_{i}\qty[ \hat{H} - E ] 
% \beta^{\dagger}_{\mu}\beta^{\dagger}_{\nu}}{\Phi}$ and 
% $B_{ij\mu\nu} = \ev{\beta_{j}\beta_{i}\beta_{\nu}\beta_{\mu}}{\Phi}$. 
% Tong: Commented equations for A and B are not correct.
Solving \eqref{eq:LR_matrix} directly can become computationally prohibitive in the 
case of heavy deformed nuclei where the number of relevant quasiparticles can 
be of the order of $10^3$, leading $A$ and $B$ to be complex-valued dense 
matrices of size of about $10^6$.

%%%%%%%%%%%%%%%%%%%%%%%%%%%%%%%%%%%%%%%%%%%%%%%%%%%%%%%%%%%%%%%%%%%%%%%%%%%%%%%

\subsubsection{Finite-Amplitude Method}

The finite-amplitude method (FAM) provides an alternative way to determine the 
amplitudes $X$ and $Y$ characterizing the perturbed density and scales much 
more favorably than the direct matrix approach \cite{nakatsukasa2007finite,
avogadro2011finite}. The FAM yields a system of coupled equations
\begin{align}
\qty( E_{\mu} + E_{\nu} - \omega ) X_{\mu\nu}(\omega) + \delta H^{20}_{\mu\nu}(\omega) & = -F^{20}_{\mu\nu}(\omega)\, ,
\label{eq:LR_FAM_1}
\\
\qty( E_{\mu} + E_{\nu} + \omega ) Y_{\mu\nu}(\omega) + \delta H^{02}_{\mu\nu}(\omega) & = -F^{02}_{\mu\nu}(\omega)\, , 
\label{eq:LR_FAM_2}
\end{align}
where $\delta H^{20}$ and $\delta H^{02}$ are functionals of the 
perturbed densities 
\begin{align}
\delta\rho       & = UXV^{T} + V^{*}Y^{T}U^{\dagger} \, , \label{eq:d_rho}\\
\delta\kappa_+     & = UXU^{T} + V^{*}Y^{T}V^{\dagger} \, , \\
\delta\kappa_- & = V^{*}X^{\dagger}V^{\dagger} + UY^{*}U^{T} \, ,
\end{align}
given by the Bogoliubov transformation of $\delta\mathcal{R}(\omega)$ as
\begin{equation}
\mqty(
     \delta\rho & \delta\kappa_+ \\
     \delta\kappa_-^\dagger & -\delta\rho^*
)
= 
\mqty(
     U & V^* \\
     V & U^*
)
\mqty(  0 & X \\
Y^T & 0
)
\mqty(
     U^\dagger & V^\dagger \\
     V^T & U^T
)\, .
\end{equation}
Given the initial HFB solution characterized by the matrices $U$ and $V$ and 
the matrix elements $F^{20}$ and $F^{02}$ of the perturbation operator $\hat{F}$, 
equations \eqref{eq:LR_FAM_1}-\eqref{eq:LR_FAM_2} can be solved iteratively.

After the $X$ and $Y$ amplitudes have been determined, the actual excitation  
strength can be readily obtained as 
\begin{equation}
\frac{dB(\omega)}{d\omega} = -\frac{1}{\pi} \text{Im}\, S(\omega) \, ,
\label{eq:response}
\end{equation}
with the response function $S(\omega)$ given by
\begin{equation}
S(\omega) = \sum_{\mu < \nu} \qty[ 
F^{20*}_{\mu\nu}X_{\mu\nu}(\omega) + F^{02*}_{\mu\nu}Y_{\mu\nu}(\omega)
] \, .
\label{eq:S}
\end{equation}
In practice, calculations are performed at complex values of the frequency 
$\omega$ as $\omega \rightarrow \omega + i\tfrac{\Gamma}{2}$, 
where $\Gamma$ is the smearing width. 
This is not only mathematically motivated by the need to avoid the 
poles of the response function on the real axis, but also corresponds, physically, to folding the 
solutions to \eqref{eq:LR_matrix} with a Lorentzian distribution. 
In the following we will choose $\Gamma = 1.0$ MeV if it is not explicitly specified. 

%%%%%%%%%%%%%%%%%%%%%%%%%%%%%%%%%%%%%%%%%%%%%%%%%%%%%%%%%%%%%%%%%%%%%%%%%%%%%%%

\subsubsection{Quasiparticle Cutoff}
\label{subsubsec:cutoff}

Additional complications emerge when performing practical calculations with 
zero-range pairing forces because of the ultra-violet divergence of the pairing 
tensor \cite{bulgac2002renormalization,borycki2006pairing,schunck2019energy}. 
At the HFB level, all quasiparticles $\mu$ such that $E_{\mu} > E_{\rm cut}$ 
are discarded. Depending on the value of the actual cutoff energy and of the 
parameters of the HO basis, this implies that the $U$ and $V$ matrices may not 
be square but of dimension $(N,N_{\rm qp})$ with $N$ the size of the basis and 
$N_{\rm qp} \leq N$ the actual number of active quasiparticles. In such a case, 
the one-body density matrix reads
\begin{equation}
\rho_{ij} = \sum_{\mu \leq N_{\rm qp}} V_{i\mu}V^{*}_{\mu j} \, .
\label{eq:rho}
\end{equation}
How to handle quasiparticle truncation at the FAM level is less clear. Writing 
\eqref{eq:d_rho} explicitly, we have for example
\begin{equation}
\delta\rho_{ij} = 
\sum_{\mu,\nu} U_{i\mu}X_{\mu\nu}V_{\nu j} 
+ 
\sum_{\mu,\nu} V^{*}_{i\mu}Y_{\mu\nu}U^{*}_{\nu j} \, .
\label{eq:drho}
\end{equation}
The question here is whether summations over both $\mu$ and $\nu$ should be 
truncated up to $N_{\rm qp}$, only one of them, or none. Some insight can be 
gained by examining the time-dependent evolution of the quasiparticle operator 
$\beta_{\mu}(t)$ of TDHFB,
\begin{equation}
\beta_{\mu}(t) = \qty[ \beta_{\mu} + \delta\beta_{\mu}(t) ]e^{iE_{\mu}t}\, ,
\qquad
\delta\beta_{\mu}(t) = \sum_{\nu} \qty[ 
X_{\nu\mu}(\omega)e^{-i\omega t} 
+ 
Y^{*}_{\nu\mu}(\omega)e^{i\omega t}] \beta^{\dagger}_{\nu} \, .
\end{equation}
Quasiparticle index $\mu$ should obey $\mu \leq N_{\rm qp}$ to be consistent 
with the definition of the one-body density matrix \eqref{eq:rho}. However, 
the expression for $\delta\beta_{\mu}(t)$ is not informative about any 
additional truncation on $\nu$. Therefore, one may be allowed to let the row 
indices of the matrices $X$ and $Y$ span the whole quasiparticle space while 
their column indices must be truncated. We realize that this is an {\em a minima} 
argument that should call for further clarifications, perhaps by extending the 
analysis the behavior of the HFB densities $\rho$ and $\kappa$ in homogeneous 
nuclear matter at the limit of large momenta; see discussion in Chapter 4 of 
\cite{schunck2019energy}. In the rest of this proceeding, we will adopt the 
``1/2'' truncation (only column indices are truncated).

%%%%%%%%%%%%%%%%%%%%%%%%%%%%%%%%%%%%%%%%%%%%%%%%%%%%%%%%%%%%%%%%%%%%%%%%%%%%%%%
%%%%%%%%%%%%%%%%%%%%%%%%%%%%%%%%%%%%%%%%%%%%%%%%%%%%%%%%%%%%%%%%%%%%%%%%%%%%%%%
%%%%%%%%%%%%%%%%%%%%%%%%%%%%%%%%%%%%%%%%%%%%%%%%%%%%%%%%%%%%%%%%%%%%%%%%%%%%%%%
%%%%%%%%%%%%%%%%%%%%%%%%%%%%%%%%%%%%%%%%%%%%%%%%%%%%%%%%%%%%%%%%%%%%%%%%%%%%%%%

\section{Benchmark Results}
\label{sec:results}

In this section, we present a series of calculations aimed at benchmarking the 
convergence of FAM calculations of the response function as a function of 
different numerical input parameters. All calculations were performed in the 
$^{240}$Pu nucleus with the SLy4 parameterization of the Skyrme force 
\cite{chabanat1997skyrme}. In the pairing channel, we use 
$V_{0}^{(n)} = -219.973$ MeV and $V_{0}^{(p)}= -285.524$ MeV as in 
\cite{navarroperez2022controllinga}. HFB calculations were performed with the 
code HFBTHO \cite{marevic2022axiallydeformed} (axial symmetry) using a 
quadrature grid of $N_{\rm GH} = N_{\rm GL} = 40$ Gauss-Hermite and 
Gauss-Laguerre integration points and $N_{\rm Leg} = 80$ Gauss-Legendre 
integration points \cite{stoitsov2013axially}.

In this work, we considered both electric ($EL$) and magnetic ($ML$) operators. 
We looked at the convergence patterns of both isoscalar and isovector operators 
and, for each operator, at the $K=0$ and $K>0$ components. Results 
were very similar in all cases, so we will only show results for the isoscalar, 
electric quadrupole ($E2$) transition with $K=0$ defined by the operator
\begin{equation}
\hat{F} = e\sum_{i=1}^{A} r_{i}^{2} Y_{20}(\theta_i, \varphi_i)\, , 
\end{equation}
where the $Y_{lm}(\theta,\varphi)$ are spherical harmonics 
\cite{varshalovich1988quantum} and we took $e$ as the regular electric charge.

%%%%%%%%%%%%%%%%%%%%%%%%%%%%%%%%%%%%%%%%%%%%%%%%%%%%%%%%%%%%%%%%%%%%%%%%%%%%%%%
%%%%%%%%%%%%%%%%%%%%%%%%%%%%%%%%%%%%%%%%%%%%%%%%%%%%%%%%%%%%%%%%%%%%%%%%%%%%%%%

\subsection{Convergence of the Response Function with Respect to the HO Basis}
\label{subsec:conv_basis}

Both the HFB and FAM equations are solved in the harmonic oscillator (HO) 
basis. Basis functions are characterized by an oscillator length $b_0$, which 
is equivalent to an oscillator frequency $\omega_0 = \hbar/(m b_0^2)$. 
In realistic calculations of deformed nuclei, the basis itself is deformed, or 
stretched. For axially-deformed basis, this means that the two frequencies 
$\omega_z$ and $\omega_{\perp}$ are different, leading to different oscillator 
lengths $b_z \neq b_{\perp}$. In HFBTHO, the ratios of the two frequencies is 
related to the basis deformation parameter $\beta_2$; see 
\cite{stoitsov2013axially} for details. The oscillator is thus characterized by 
two parameters, $b_0$ and $\beta_2$. The total number of basis functions is 
determined by the number of shells $N_\mathrm{sh}$. 

\begin{figure}[!h]
\centering
\includegraphics[width=0.475\textwidth,clip]{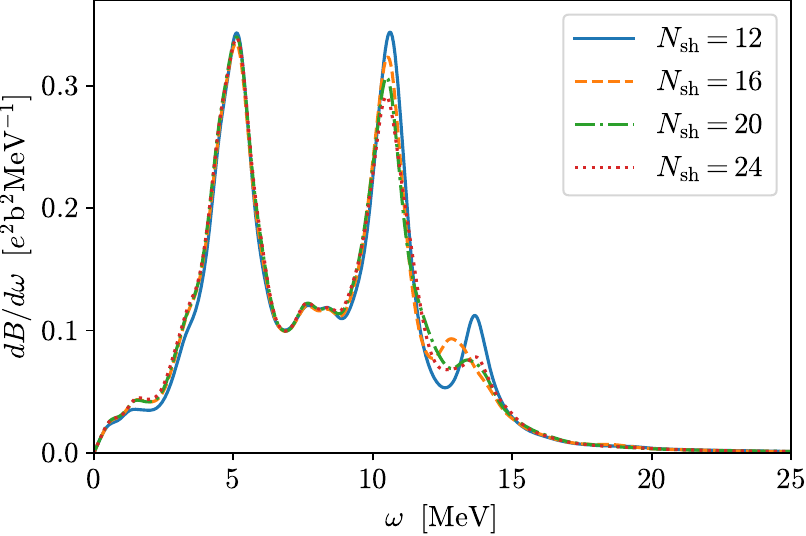}
\includegraphics[width=0.49\textwidth,clip]{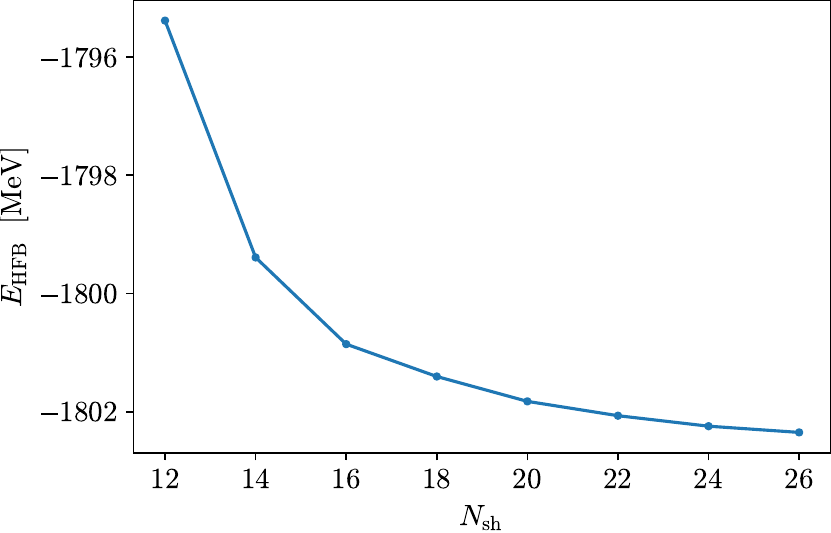}
\caption{Left: Isoscalar electric quadrupole excitation strength $dB(\omega)/d\omega$ 
in $^{240}$Pu as a function of the excitation energy $\omega$,  for different values of the total 
number of shells in the basis. Right: Variation of the HFB energy across the same range 
of oscillator shells.
}
\label{fig:var_N}
\end{figure}

In the left panel of Fig.\ref{fig:var_N}, we show how the isoscalar, electric 
quadrupole excitation strength changes with the size of the basis as determined 
by the total number of shells $N_\mathrm{sh}$. For each calculation, the oscillator 
length is $b_0 = 2.3$ fm and the basis is spherical, $\beta_2 = 0$. It is 
somewhat surprising to see that the first peak at $\omega \approx 5$ MeV 
is hardly changed, the height of the second peak at $\omega \approx 10$ MeV varies by less than 15\%, 
and most of the variability of the results is in the small third peak at around $\omega \approx 12$ MeV. 
This rather weak dependency of the results with the number of shells should be 
contrasted with the rather strong dependency of the HFB energy; see the right panel 
of Fig.\ref{fig:var_N}.

\begin{figure}[!h]
\centering
\includegraphics[width=0.47\textwidth,clip]{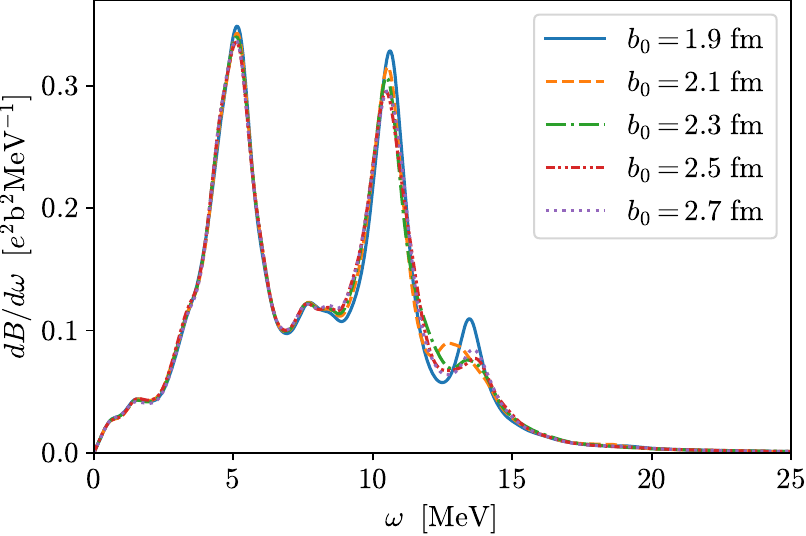}
\includegraphics[width=0.49\textwidth,clip]{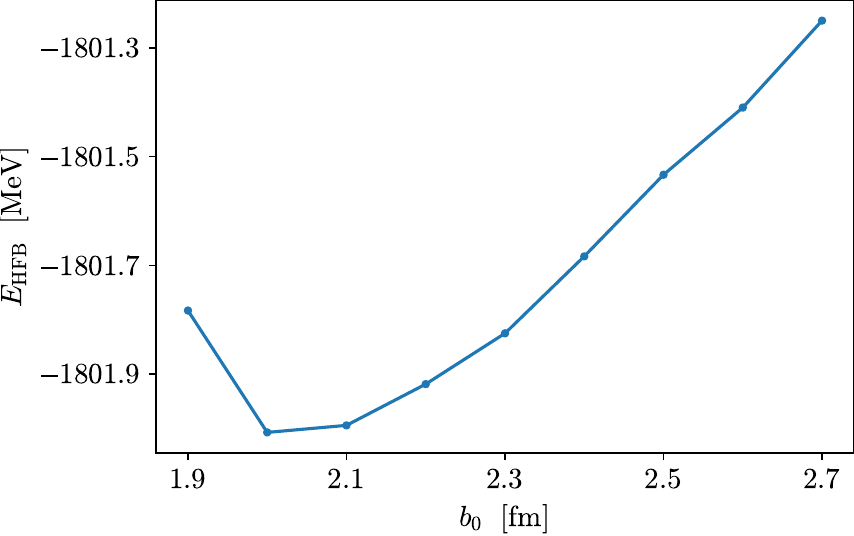}
\caption{Left: Isoscalar electric quadrupole excitation strength $dB(\omega)/d\omega$ 
in $^{240}$Pu as a function of the excitation energy $\omega$, for different values of the oscillator 
length (in fm). Right: Variation of the HFB energy across the same range of oscillator 
lengths. 
}
\label{fig:var_b}
\end{figure}

In the left panel of Fig.\ref{fig:var_b}, we show a similar study for the 
variations of the response with the oscillator length $b_0$. Here, in each 
calculation the number of shells is $N_\mathrm{sh} = 20$ and the basis is also spherical. 
Although there are some small differences in the height of each of the peaks, 
the response function is pretty insensitive to the value of $b_0$. The right 
panel of the figure shows that the variations of the HFB energy with $b_0$ do 
not exceed 700 keV at $N_\mathrm{sh}= 20$. Similar conclusions apply when varying the 
deformation $\beta_2$ of the basis.

%%%%%%%%%%%%%%%%%%%%%%%%%%%%%%%%%%%%%%%%%%%%%%%%%%%%%%%%%%%%%%%%%%%%%%%%%%%%%%%
%%%%%%%%%%%%%%%%%%%%%%%%%%%%%%%%%%%%%%%%%%%%%%%%%%%%%%%%%%%%%%%%%%%%%%%%%%%%%%%

\subsection{Impact of Quasiparticle Cutoff}
\label{subsec:conv_ecut}

\begin{figure}[!h]
\centering
\includegraphics[width=0.48\textwidth,clip]{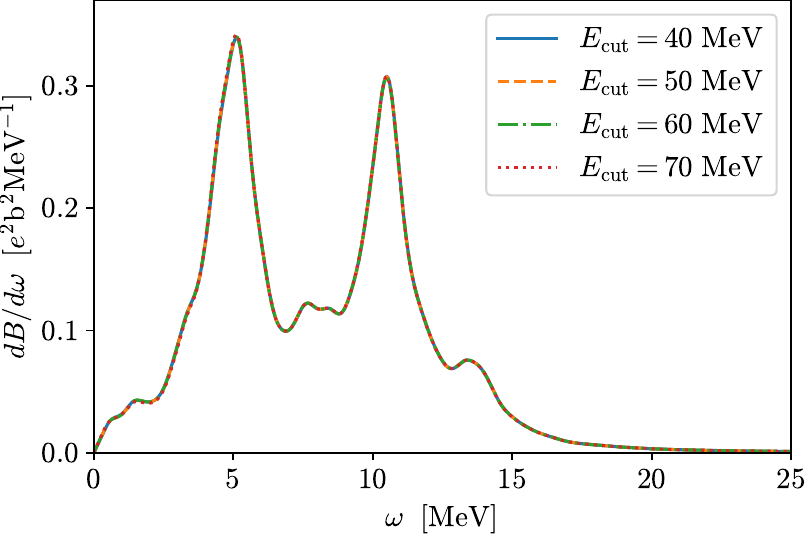}
\includegraphics[width=0.48\textwidth,clip]{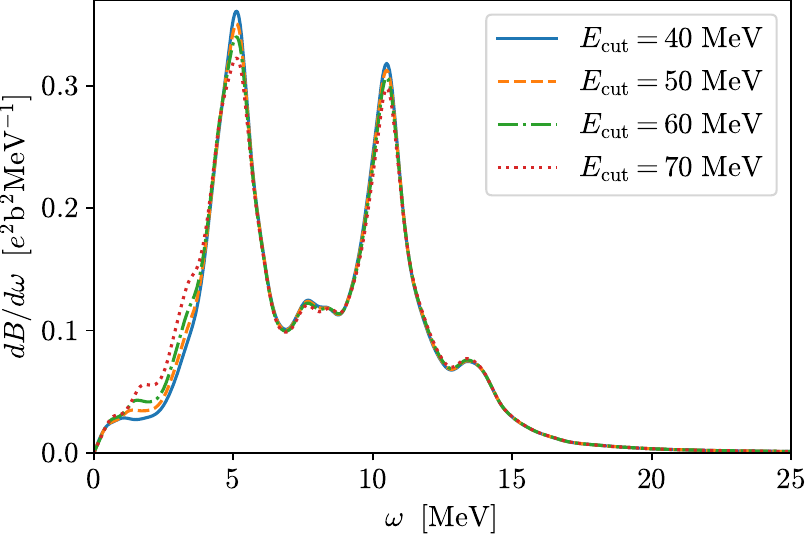}
\caption{Left: Isoscalar electric quadrupole excitation strength $dB(\omega)/d\omega$ 
in $^{240}$Pu as a function of the excitation energy $\omega$, for different quasiparticle cutoff energies. 
For each $E_{\rm cut}$, the strengths $V_{0}^{(n)}$ and 
$V_{0}^{(p)}$ of the pairing force have been refitted to keep the 
pairing gaps of $^{232}$Th at their reference values. Right: Same as in the left panel, only a 
single HFB solution with $E_{\rm cut} = 60$ MeV is used for all four FAM 
calculations; see text for additional details.}
\label{fig:var_Ecut}
\end{figure}

As mentioned earlier, HFB and QRPA calculations with zero-range pairing forces 
require introducing a cutoff on the number of quasiparticles used to define the 
densities \eqref{eq:rho} and \eqref{eq:drho}. In Fig.\ref{fig:var_Ecut}, we 
illustrate the dependence of the response function on the choice of the cutoff. 

Given a cutoff value $E_{\rm cut}$, a consistent calculation requires first 
calibrating the strengths $V_{0}^{(n)}$ and $V_{0}^{(p)}$ of the pairing force 
on some experimental data. Here we perform this calibration at the HFB level by 
employing the same prescription as in \cite{navarroperez2022controllinga}. In 
practice, for each $E_{\rm cut}$ we refitted the pairing strengths so that the 
HFB pairing gaps match the three-point odd-even mass differences $\Delta^{(3)}$ in $^{232}$Th. 
After this calibration, the pairing channel of the FAM calculations is well 
defined and we can compute the perturbed densities \eqref{eq:drho} according to 
the prescription discussed in Section \ref{subsubsec:cutoff}. 

In the left panel of Fig.~\ref{fig:var_Ecut}, we show results for the same 
isoscalar electric quadrupole excitation strength as before for four values of the cutoff. 
Not surprisingly, calculations are undistinguishable from one another since 
recalibrating the pairing strengths imply that the underlying HFB solutions are 
nearly identical by construction. For completeness, we also show in the right 
panel the non-consistent case where the cutoff of the HFB solution is fixed, 
here at $E_{\rm cut} = 60$ MeV, while the cutoff of the FAM calculation 
\eqref{eq:drho} itself varies. In other words a single identical HFB 
calculation is used as the basis for all FAM calculations. Even in this not very 
realistic scenario, the response depends only weakly on the cutoff.

%%%%%%%%%%%%%%%%%%%%%%%%%%%%%%%%%%%%%%%%%%%%%%%%%%%%%%%%%%%%%%%%%%%%%%%%%%%%%%%
%%%%%%%%%%%%%%%%%%%%%%%%%%%%%%%%%%%%%%%%%%%%%%%%%%%%%%%%%%%%%%%%%%%%%%%%%%%%%%

\subsection{Dependence of the Response Function on the Smearing Width}
\label{subsec:conv_fam}

Finally, we discuss the dependency of the results on the smearing width 
$\Gamma$ of the FAM response. As mentioned earlier, this parameter is 
introduced because the response function \eqref{eq:response} has poles 
corresponding to QRPA eigenmodes: computing the response at 
complex $\omega$ values makes calculations more efficient 
\cite{nakatsukasa2007finite}. At the same time, it is equivalent to computing 
the response \eqref{eq:S} by summing explicitly over the solutions of 
\eqref{eq:LR_matrix} and folding the result with a Lorentzian distribution of 
width $\Gamma$. The net result is that $\Gamma$ becomes a parameter of the 
calculation and its choice will have an impact on the results.

\begin{figure}[!h]
\centering
\includegraphics[width=0.7\textwidth,clip]{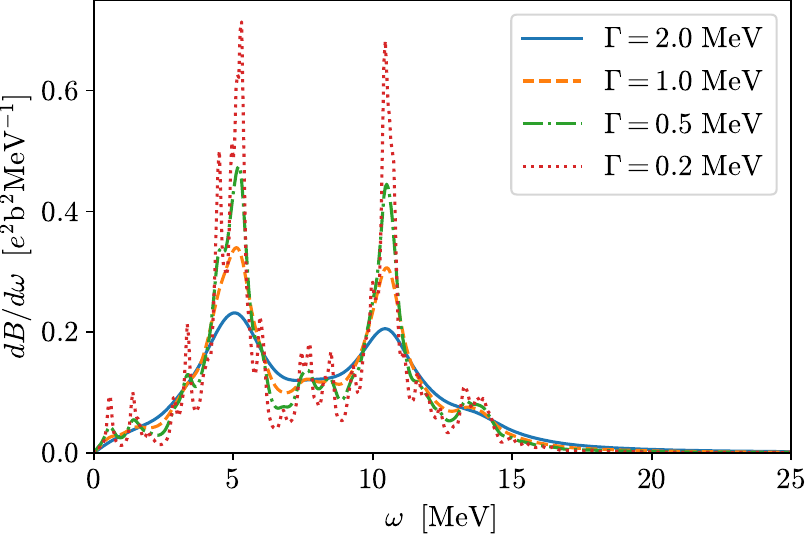}
\caption{Isoscalar electric quadrupole excitation strength $dB(\omega)/d\omega$ 
in $^{240}$Pu as a function of the excitation energy $\omega$, for different smearing widths $\Gamma$.}
\label{fig:var_FAM}
\end{figure}

From a computational perspective, large values of $\Gamma$ make calculations 
faster and more stable. From a physics standpoint, semi-classical models of the 
giant dipole response are typically based on a fit of the photoabsorption 
cross section with a Lorentzian shape \cite{bohr1998nucleara}. Surveys of 
experimental results suggest a rather large variation of $\Gamma$ with the mass 
number $A$ of the nucleus \cite{goriely2019reference}. In Fig.~\ref{fig:var_FAM} 
we show how the isoscalar electric quadrupole excitation strength changes with $\Gamma$. As 
expected, larger values of $\Gamma$ smooth out all the features of the response 
while smaller values make its fine structure more visible. Out of all the input 
parameters we considered in this work, $\Gamma$ is by far the one with the 
largest impact on the results.

%%%%%%%%%%%%%%%%%%%%%%%%%%%%%%%%%%%%%%%%%%%%%%%%%%%%%%%%%%%%%%%%%%%%%%%%%%%%%%%
%%%%%%%%%%%%%%%%%%%%%%%%%%%%%%%%%%%%%%%%%%%%%%%%%%%%%%%%%%%%%%%%%%%%%%%%%%%%%%%
%%%%%%%%%%%%%%%%%%%%%%%%%%%%%%%%%%%%%%%%%%%%%%%%%%%%%%%%%%%%%%%%%%%%%%%%%%%%%%%
%%%%%%%%%%%%%%%%%%%%%%%%%%%%%%%%%%%%%%%%%%%%%%%%%%%%%%%%%%%%%%%%%%%%%%%%%%%%%%%

\section{Conclusions}
\label{sec:conclusion}

We have investigated the convergence patterns of the electromagnetic response 
function as a function of the numerical parameters controlling the calculation. 
Overall, the dependency of the results HO basis parameters is weaker than that of 
the underlying HFB calculations themselves. The response function is also 
insensitive to the quasiparticle cutoff -- when the latter is implemented 
consistently at both HFB and FAM levels. Only the smearing width $\Gamma$ used 
to define the complex frequency $\omega$ of the response has a visible impact 
on the results.

%%%%%%%%%%%%%%%%%%%%%%%%%%%%%%%%%%%%%%%%%%%%%%%%%%%%%%%%%%%%%%%%%%%%%%%%%%%%%%%
%%%%%%%%%%%%%%%%%%%%%%%%%%%%%%%%%%%%%%%%%%%%%%%%%%%%%%%%%%%%%%%%%%%%%%%%%%%%%%%
%%%%%%%%%%%%%%%%%%%%%%%%%%%%%%%%%%%%%%%%%%%%%%%%%%%%%%%%%%%%%%%%%%%%%%%%%%%%%%%
%%%%%%%%%%%%%%%%%%%%%%%%%%%%%%%%%%%%%%%%%%%%%%%%%%%%%%%%%%%%%%%%%%%%%%%%%%%%%%%

\section*{Acknowledgments}
This work was performed under the auspices of the U.S.\ Department of Energy by
Lawrence Livermore National Laboratory under Contract DE-AC52-07NA27344. 
Computing support came from the Lawrence Livermore National Laboratory (LLNL)
Institutional Computing Grand Challenge program.

\bibliography{zotero_output,books}

\providecommand{\noopsort}[1]{}
\begin{thebibliography}{28}

\bibitem{goriely2019reference}
S.~Goriely, P.~Dimitriou, M.~Wiedeking, T.~Belgya, R.~Firestone, J.~Kopecky,
  M.~Krti{\v c}ka, V.~Plujko, R.~Schwengner, S.~Siem et~al., Eur. Phys. J. A
  \textbf{55}, 172 (2019)

\bibitem{goriely2002largescale}
S.~Goriely, E.~Khan, Nucl. Phys. A \textbf{706}, 217 (2002)

\bibitem{goriely2004microscopic}
S.~Goriely, E.~Khan, M.~Samyn, Nucl. Phys. A \textbf{739}, 331 (2004)

\bibitem{daoutidis2012largescale}
I.~Daoutidis, S.~Goriely, Phys. Rev. C \textbf{86}, 034328 (2012)

\bibitem{martini2016largescale}
M.~Martini, S.~P{\'e}ru, S.~Hilaire, S.~Goriely, F.~Lechaftois, Phys. Rev. C
  \textbf{94}, 014304 (2016)

\bibitem{goriely2018gognyhfb}
S.~Goriely, S.~Hilaire, S.~P{\'e}ru, K.~Sieja, Phys. Rev. C \textbf{98}, 014327
  (2018)

\bibitem{nakatsukasa2007finite}
T.~Nakatsukasa, T.~Inakura, K.~Yabana, Phys. Rev. C \textbf{76}, 024318 (2007)

\bibitem{avogadro2011finite}
P.~Avogadro, T.~Nakatsukasa, Phys. Rev. C \textbf{84}, 014314 (2011)

\bibitem{inakura2009selfconsistent}
T.~Inakura, T.~Nakatsukasa, K.~Yabana, Phys. Rev. C \textbf{80}, 044301 (2009)

\bibitem{stoitsov2011monopole}
M.~Stoitsov, M.~Kortelainen, T.~Nakatsukasa, C.~Losa, W.~Nazarewicz, Phys. Rev.
  C \textbf{84}, 041305 (2011)

\bibitem{kortelainen2015multipole}
M.~Kortelainen, N.~Hinohara, W.~Nazarewicz, Phys. Rev. C \textbf{92}, 051302(R)
  (2015)

\bibitem{oishi2016finite}
T.~Oishi, M.~Kortelainen, N.~Hinohara, Phys. Rev. C \textbf{93}, 034329 (2016)

\bibitem{petrik2018thoulessvalatin}
K.~Petr{\'i}k, M.~Kortelainen, Phys. Rev. C \textbf{97}, 034321 (2018)

\bibitem{washiyama2021finiteamplitude}
K.~Washiyama, N.~Hinohara, T.~Nakatsukasa, Phys. Rev. C \textbf{103}, 014306
  (2021)

\bibitem{niksic2013implementation}
T.~Nik{\v s}i{\'c}, N.~Kralj, T.~Tuti{\v s}, D.~Vretenar, P.~Ring, Phys. Rev. C
  \textbf{88}, 044327 (2013)

\bibitem{valatin1961generalized}
J.G. Valatin, Phys. Rev. \textbf{122}, 1012 (1961)

\bibitem{mang1975selfconsistent}
H.J. Mang, Phys. Rep. \textbf{18}, 325 (1975)

\bibitem{bender2003selfconsistent}
M.~Bender, P.H. Heenen, P.G. Reinhard, Rev. Mod. Phys. \textbf{75}, 121 (2003)

\bibitem{ring2004nuclear}
P.~Ring, P.~Schuck, \emph{The Nuclear Many-Body Problem}, Texts and Monographs
  in Physics (Springer, 2004)

\bibitem{schunck2019energy}
N.~Schunck, \emph{Energy Density Functional Methods for Atomic Nuclei.}, {{IOP
  Expanding Physics}} ({IOP Publishing}, {Bristol, UK}, 2019), ISBN
  978-0-7503-1423-7,
  \urlstyle{tt}\url{https://doi.org/10.1088/2053-2563/aae0ed}

\bibitem{bulgac2002renormalization}
A.~Bulgac, Y.~Yu, Phys. Rev. Lett. \textbf{88}, 042504 (2002)

\bibitem{borycki2006pairing}
P.~Borycki, J.~Dobaczewski, W.~Nazarewicz, M.~Stoitsov, Phys. Rev. C
  \textbf{73}, 044319 (2006)

\bibitem{chabanat1997skyrme}
E.~Chabanat, P.~Bonche, P.~Haensel, J.~Meyer, R.~Schaeffer, Nucl. Phys. A
  \textbf{627}, 710 (1997)

\bibitem{navarroperez2022controllinga}
R.~Navarro~P{\'e}rez, N.~Schunck, Phys. Lett. B \textbf{833}, 137336 (2022)

\bibitem{marevic2022axiallydeformed}
P.~Marevi{\'c}, N.~Schunck, E.M. Ney, R.~Navarro~P{\'e}rez, M.~Verriere,
  J.~O'Neal, Comput. Phys. Commun. \textbf{276}, 108367 (2022)

\bibitem{stoitsov2013axially}
M.~Stoitsov, N.~Schunck, M.~Kortelainen, N.~Michel, H.~Nam, E.~Olsen,
  J.~Sarich, S.~Wild, Comput. Phys. Commun. \textbf{184}, 1592 (2013)

\bibitem{varshalovich1988quantum}
D.~Varshalovich, A.~Moskalev, V.~Khersonskii, \emph{Quantum Theory of Angular
  Momentum} (World Scientific, Singapore, 1988)

\bibitem{bohr1998nucleara}
A.~Bohr, B.~Mottelson, \emph{Nuclear Structure}, Vol. II, Nuclear Deformations
  (World Scientific, 1998)

\end{thebibliography}

\end{document}